\documentclass[11pt]{amsart}

\usepackage[T1]{fontenc}
\usepackage[utf8]{inputenc}
\usepackage{lmodern}
\usepackage{amsmath,amssymb,amsthm,mathtools}
\usepackage{enumitem}
\usepackage{booktabs}
\usepackage[margin=1.15in]{geometry}
\usepackage{xcolor}
\usepackage{mdframed}
\usepackage{hyperref}
\usepackage{tikz-cd}

\hypersetup{
  colorlinks=true, linkcolor=blue!60!black, citecolor=blue!60!black, urlcolor=blue!60!black }

\theoremstyle{definition}
\newtheorem{definition}{Definition}[section]

\theoremstyle{plain}
\newtheorem{theorem}[definition]{Theorem}

\newtheorem{proposition}[definition]{Proposition}

\newtheorem{remark}[definition]{Remark}

\newcommand{\Lag}{\mathrm{Lag}}
\newcommand{\PreqLag}{\mathrm{PreqLag}}

\newcommand{\Map}{\operatorname{Map}}

\newcommand{\pt}{\operatorname{pt}}

\newcommand{\Tcot}[2]{T^*[#1]#2}
\newcommand{\qstack}[2]{[#1/#2]}
\newcommand{\Qstack}[2]{[#1/\!/#2]}

\title[Classical Symmetry Topological Field Theories]{Classical Symmetry TFTs for Continuous Symmetries via Higher Symplectic Geometry}
\author{Hao Xu}
\date{\today}

\begin{document}

\begin{abstract}
We propose a shifted-symplectic formulation of a classical continuous analogue of the symmetry TFT paradigm. Let $G$ be an algebraic or Lie group acting by topological defects on an $n$-dimensional classical topological sigma model with target an $(n-1)$-shifted symplectic derived stack $(X,\omega)$ via the AKSZ construction. We argue that the corresponding $(n+1)$-dimensional bulk theory should be the AKSZ theory with target the shifted cotangent stack $\Tcot{n}{\mathrm B G}$, equivalently the $(n+1)$-dimensional BF theory for $G$. We characterize the Dirichlet and Neumann boundary conditions, and more general topological boundaries, in terms of shifted Lagrangians in $\Tcot{n}{\mathrm B G}$. We realize the gauging of the $G$-symmetry in the original theory as inserting a topological domain wall between the corresponding topological boundaries in the BF bulk, and introduce the notion of Hamiltonian reduction, syplectic reduction, and Lagrangian reduction in the shifted symplectic setting. We also discuss prequantum refinements of continuous SymTFTs. In this refinement, higher gerbes on $\mathrm B G$ encode classical analogues of 't Hooft anomaly data by decorating the shifted cotangent bulk and its Lagrangian boundary conditions. Finally, in dimension three we compare the infinitesimal BF model $\mathrm B(\mathfrak g\ltimes\mathfrak g^\vee)$ with the factorizable double $\mathrm B(\mathfrak g\oplus \mathfrak g)$. The resulting topological boundaries are described by Lagrangian Lie subalgebras, and the factorizable case relates the SymTFT dictionary to $r$-matrices and Belavin--Drinfeld data.
\end{abstract}

\maketitle
\tableofcontents

\section{Introduction}

The symmetry topological field theory (SymTFT) paradigm packages the symmetry of a $d$-dimensional quantum field theory into a $(d+1)$-dimensional bulk theory placed on an interval \cite{KW14,KLWZZ20,FMT22}. One boundary is topological and encodes the abstract symmetry data. The other is physical and encodes the particular theory. Compactifying the interval recovers the original theory, and gauging is implemented by changing the topological boundary while leaving the physical boundary fixed. Moreover, one can classify topological defects in the original theory by the topological defects in the symmetry TFT, which are often easier to analyze.

This paper proposes a classical analogue for continuous symmetries. The ambient language is the AKSZ construction in derived algebraic geometry as developed by Calaque, Haugseng and Scheimbauer \cite{CHS25}, building on shifted symplectic geometry \cite{PTVV13} and Lagrangian structures on mapping stacks \cite{Calaque15}. In this setting the target of classical field theories are organized as iterative shifted Lagrangian correspondences. This is exactly the structure needed for a classical version of the symmetry TFT sandwich.

The basic proposal is simple. If $G$ acts on an $n$-shifted symplectic stack $(X,\omega)$ by shifted Lagrangian self-correspondences, then the natural candidate for the continuous symmetry TFT bulk is not classical Chern--Simons theory, but rather the topological sigma model on the shifted cotangent stack $\Tcot{n}{\mathrm B G}$, equipped with its canonical $n$-shifted symplectic form \cite{Calaque19}. This is the classical BF theory for $G$. Thus the slogan is
\begin{center}
\fbox{$\displaystyle\begin{aligned}
  \text{classical continuous SymTFT}
  &=
  \text{BF bulk}
  +
  \text{Dirichlet boundary} \\
  &\quad+
  \text{physical boundary}.
\end{aligned}$}
\end{center}

In general, a $G$-action on $X$ can be \emph{anomalous} on the associated AKSZ theory, for example, twisted by a higher gerbe on the classifying stack $\mathrm{B} G$; this anomaly can be encoded in a twisted BF bulk. More precisely, the untwisted theory below is a \emph{classical} TFT valued in $\Lag_n$, while an anomaly class belongs to the \emph{prequantum} refinement: one decorates the shifted symplectic target and its Lagrangian boundary conditions by compatible higher gerbes with connection. Forgetting the gerbes returns the classical sandwich, but keeping them turns higher gerbes into invertible objects which may be stacked with the BF bulk or with a boundary condition.

\subsection{Main Principle}

The paper is organized around the following principle.

\begin{proposition}[Shifted Symplectic Reduction]\label{conj:main}
Let $(X,\omega)$ be an $(n-1)$-shifted symplectic derived stack and let $G$ act on $X$ by symplectomorphisms. 

\begin{enumerate}[label=(\roman*)]
  \item The shifted conormal of the quotient stack $\qstack{X}{G}$ provides a shifted Lagrangian of the shifted cotangent stack: \[ N^* [n] (\qstack{X}{G} \to \mathrm{B}G) \to \Tcot{n}{\mathrm B G},\]
  
  \item The quotient stack $\qstack{X}{G}$ is lifted to an $n$-shifted Lagrangian in $\Tcot{n}{\mathrm B G}$ if and only if $X$ admits a shifted moment map, or equivalently, the $G$-action on $X$ is Hamiltonian;
  
  \item For any closed subgroup $H\leq G$, there is an $n$-shifted Lagrangian given by the shifted conormal stack: \[N^* [n] (\mathrm{B}H \to \mathrm{B}G) \longrightarrow \Tcot{n}{\mathrm B G},\] where for $H = *$ this is the topological Dirichlet boundary $\mathfrak{g}^\vee [n-1]$ , and for $H = G$ this is the topological Neumann boundary $\mathrm{B}G$;
  
  \item The topological gauging from symmetry $G$ to symmetry $H$ is implemented by the topological domain wall between the corresponding Lagrangians:
  \[\begin{tikzcd}
    {(\mathfrak{g}/\mathfrak{h})^\vee [n-1]}
      \arrow[d]
      \arrow[r]
    & {N^* [n] (\mathrm{B}H \to \mathrm{B}G)}
      \arrow[d]
      \arrow[r]
    & {\pt}
    \\ {\mathfrak{g}^\vee [n-1]}
      \arrow[d]
      \arrow[r]
    & {\Tcot{n}{\mathrm B G}}
    & {}
    \\ {\pt}
    & {}
    & {}
  \end{tikzcd};\]
  
  \item The Hamiltonian reduction of $X$ by $H$ is given by the derived intersection:
  \[\begin{tikzcd}
    {X}
    & {}
    & {\qstack{X}{G} \times_{\Tcot{n}{\mathrm B G}} \mathfrak{g}^\vee [n-1]}
      \arrow[ll,"\sim"']
    \\ {\mu_H^{-1}(0)}
      \arrow[u]
      \arrow[d]
    & {}
    & {\qstack{X}{G} \times_{\Tcot{n}{\mathrm B G}}( \mathfrak{g}/\mathfrak{h})^\vee [n-1]}
      \arrow[u]
      \arrow[d]
      \arrow[ll,"\sim"]
    \\ {\Qstack{X}{H}}
    & {}
    & {\qstack{X}{G} \times_{\Tcot{n}{\mathrm B G}} N^* [n] (\mathrm{B}H \to \mathrm{B}G)} 
      \arrow[ll,"\sim"]
    \end{tikzcd},\] where $\mu_H \colon X \to \mathfrak{h}^\vee [n-1]$ is the shifted moment map to $H$-symmetry.
\end{enumerate}
\end{proposition}

This statement is essentially a repackaging and extension of Anel--Calaque's results \cite{AnelCalaque21} on shifted symplectic reduction.

\subsection*{Literatures}
The subject of generalized symmetries and symmetry TFTs is a subject yet to be fully developed. We note that the following is far from a complete list of references. On the side of topological orders, Kong and Wen laid an early framework by uncovering hidden categorical structures arising from gravitational anomalies \cite{KW14}. In parallel, within quantum field theory, Gaiotto, Kapustin, Seiberg, and Willett introduced the notion of generalized global symmetries and their anomalies \cite{GKSW15}, sparking a wave of interest in the symmetry TFT paradigm \cite{KWZ,KLWZZ20, FMT22, KaidiOhmoriZheng22}.

Closer to the present setting, a number of works have studied symmetry TFTs for \emph{continuous} symmetries, where the relevant bulk theories are no longer of finite type. The string-theoretic origin of SymTFTs and their geometric engineering was developed by Apruzzi, De Luca, Gnecchi, Lo Monaco, and Sch\"afer-Nameki \cite{ApruzziEtAl21}, with an extension to non-finite (non-compact) symmetry theories \cite{ApruzziBedognaDondi24}. A SymTFT for continuous symmetries, built from $\mathrm{BF}$-type bulk actions, was proposed by Brennan and Sun \cite{BrennanSun24b}, while Antinucci and Benini analyzed anomalies and gauging of $U(1)$ symmetries from this viewpoint \cite{AntinucciBenini24}. Bonetti, Del Zotto, and Minasian gave a systematic treatment of SymTFTs for continuous non-abelian symmetries \cite{BonettiDelZottoMinasian24}, later incorporating non-linear realizations and spontaneous symmetry breaking \cite{BonettiDelZottoMinasian25}; a complementary derivation of $U(1)$ SymTFTs via descent was given by Gagliano and Garc\'ia Etxebarria \cite{GaglianoGarciaEtxebarria24}, and the framework was extended to continuous \emph{spacetime} symmetries by Apruzzi, Dondi, Garc\'ia Etxebarria, Lam, and Sch\"afer-Nameki \cite{ApruzziEtAl25}. Related developments include non-invertible $T$-duality realized through a non-compact SymTFT \cite{ArgurioEtAl24}, a SymTFT approach to flavor symmetries \cite{WangEtAl25b}, and an operator-algebraic and Hilbert-space treatment of categorical continuous symmetries \cite{WangEtAl25a, WangEtAl26a, TianJia26}. The $\mathrm{BF}$-type field content underlying many of these constructions has a long history in the study of topological observables, going back to the Wilson surface and higher knot invariants of Cattaneo and Rossi \cite{CattaneoRossi02}. Closest in spirit to the methods used here, Borsten, Kanakaris, and Kim construct SymTFTs for centre symmetries of Chern--Simons, Yang--Mills, and Einstein gravity via the Pulmann--\v{S}evera--Valach sandwich construction \cite{BorstenKanakarisKim25}, the same homotopy-theoretic machinery that underlies our shifted-symplectic treatment \cite{PullmannSeveraValach19}. While these works are largely physical and Lagrangian formalism in flavor, and our approach is instead built on shifted symplectic geometry and the AKSZ formalism, they share the common goal of organizing continuous symmetries through a topological bulk symmetry TFT.

\subsection*{Acknowledgments}
The author would like to thank his doctoral supervisor Chenchang Zhu for her continuous input of higher differential geometry and support throughout the project. The author is also grateful to Liang Kong for his insights on the physical aspects of the symmetry TFT paradigm dating back to the undergraduate days, and for his continuous encouragement and support. Thanks are also due to Miquel Cueca, Stefano Ronchi and Kalin Krishna, for sharing their expertise and updates on their on-going research projects. The author would like to thank Miquel Cueca, Alonso Perez-Lona, Fabio Apruzzi and Yinan Wang for pointing out the relevant references. The author is supported by Villum Fonden 00060714 ``\emph{Global Categorical Symmetries and Phases of Quantum Matter}''.

\section{Background}

\subsection{Shifted Symplectic AKSZ}\label{sec:background-aksz}

Pantev, Toën, Vaquié, and Vezzosi introduced $n$-shifted symplectic structures on derived Artin stacks and proved, among other things, that mapping stacks inherit shifted symplectic structures by fiber integration \cite{PTVV13}. If $F$ is $n$-shifted symplectic and $M$ is an oriented compact manifold of dimension $d$, then the derived mapping stack $\Map(M_\flat,F)$ carries a canonical $(n-d)$-shifted symplectic structure, under the usual finiteness and orientation hypotheses. Here $M_\flat$ is the de Rham stack of $M$.

Calaque extended this picture to mapping stacks with boundary conditions, producing Lagrangian structures from Lagrangian boundary data \cite{Calaque15}. The fully extended form of this story is the $(\infty,n)$-categorical AKSZ construction by Calaque--Haugseng--Scheimbauer \cite{CHS25}. Its input is an $n$-shifted symplectic stack, and its output is an oriented extended topological field theory valued in a higher category of iterated shifted Lagrangian correspondences.

We will use this construction as a black box. The only structural features needed below are:
\begin{enumerate}[label=(\roman*)]
  \item An $n$-shifted symplectic target defines a classical AKSZ-type
  field theory, 
  
  \item Lagrangian morphisms into the target define topological boundary conditions;
  
  \item Composing Lagrangian correspondences models compactification,
  cutting or gluing topological defects in AKSZ-type theories;

  \item From the perspective of geometric quantization, 0-shifted symplectic stacks are the semi-classical targets replacing Hilbert spaces, so in our physical setting, an $n$-shifted symplectic stack gives rise to a fully extended $(n+1)$-dimensional topological field theory.
\end{enumerate}

\begin{remark}[Framed and coisotropic variants]
Haugseng, Melani and Safronov construct higher categories of shifted coisotropic correspondences between derived Poisson stacks and obtain framed extended TFTs by the Cobordism Hypothesis \cite{HMS22}. This broadens the language from symplectic/Lagrangian to Poisson/coisotropic geometry. The BF target remains well behaved in this setting, since the zero section is Lagrangian and hence nondegenerate coisotropic.
\end{remark}

\subsection{The Classical 3D Chern--Simons Theory}

Classical Chern--Simons theory for a compact simple Lie group $G$ is defined by the action
\[
  S_{\mathrm{CS}}[A] = \frac{k}{4\pi}\int_M \left\langle A \wedge \mathrm{d}A \right\rangle + \tfrac{1}{3}\left\langle A \wedge [A,A] \right\rangle,
\]
where $M$ is a compact oriented $3$-manifold, $A \in \Omega^1(M,\mathfrak{g})$ is a $\mathfrak{g}$-valued connection $1$-form, and $\langle -,-\rangle$ denotes the basic invariant inner product on $\mathfrak{g}$ (normalized so that the longest root has length-squared $2$) \cite{Witten1989, Freed95}. The Euler--Lagrange equations set the curvature $F_A = \mathrm{d}A + \tfrac{1}{2}[A,A]$ to zero, so the classical solutions are flat $G$-connections. The theory is topological in the sense that no metric appears in $S_{\mathrm{CS}}$, and its gauge symmetry is the full group of $G$-valued maps on $M$.

However, the Lagrangian formulation is \emph{not} globally well-defined on a closed $3$-manifold unless the level $k$ is an integer (for simply-connected $G$ one has $\mathbb{Z} \simeq \operatorname{H}^4(\mathrm{B}G;\mathbb{Z})$), because $S_{\mathrm{CS}}$ shifts by $2\pi k$ under large gauge transformations. When $M$ has a boundary $\Sigma$, the action is no longer gauge-invariant: a gauge transformation $g\colon M\to G$ produces the boundary term $\frac{k}{4\pi}\int_\Sigma \langle g^{-1}dg \wedge A\rangle$. This anomaly is cancelled by the chiral WZW model living on $\Sigma$, which is a gapless (non-topological) $2d$ theory \cite{EMSS89, Witten1992}. 

In the AKSZ language of §\ref{sec:background-aksz}, Chern--Simons theory is the sigma model whose target is the 2-shifted symplectic stack $(\mathrm{B}G, \omega_k)$. Here $\mathrm{B}G$ is the classifying stack of $G$, and the 2-shifted symplectic form $\omega_k$ is the one induced by the $G$-invariant inner product $\langle-,-\rangle$: by the PTVV theorem \cite{PTVV13}, the level $k \in \mathrm{H}^4(\mathrm{B}G;\mathbb{Z})\simeq\mathbb{Z}$ determines a 2-shifted closed 2-form on $\mathrm{B}G$, and nondegeneracy follows from the nondegeneracy of $\langle-,-\rangle$.

\subsection{Shifted Cotangent Stacks}

For any sufficiently geometric derived stack $Y$, the shifted cotangent stack $\Tcot{n}{Y}$ carries a canonical $n$-shifted symplectic structure \cite{Calaque19}. If $f\colon Z\to Y$ is a morphism, the shifted conormal construction gives natural Lagrangian morphisms into shifted cotangent stacks.

Taking $Y = \mathrm B G$ gives the target $\Tcot{n}{\mathrm B G}$. In classical field-theoretic language, the corresponding topological sigma model is the $(n+1)$-dimensional classical BF theory. For $n=2$, and for $G$ reductive, this is the derived stack whose infinitesimal linear model is the classifying stack of the \emph{Drinfeld double} $\mathfrak g\ltimes \mathfrak g^\vee$. It is the cotangent, or doubled, version of Chern--Simons type gauge theory.

Concretely, let $M$ be a closed oriented $(n+1)$-manifold. The fields of $(n+1)$-dimensional BF theory for $G$, in a trivialization of the underlying principal $G$-bundle, are a connection $1$-form $A\in\Omega^1(M,\mathfrak g)$ and a $\mathfrak g^\vee$-valued $(n-1)$-form $B\in\Omega^{n-1}(M,\mathfrak g^\vee)$. The classical action is
\[
  S_{\mathrm{BF}}(A,B) = \int_M \langle B\wedge F_A\rangle,
\]
where $F_A = \mathrm dA + \frac{1}{2}[A\wedge A]\in\Omega^2(M,\mathfrak g)$ is the curvature and $\langle-,-\rangle\colon \mathfrak g^\vee\otimes\mathfrak g\to\mathbb R$ is the canonical pairing. The Euler--Lagrange equations are
\[
  F_A=0,\qquad \mathrm d_A^\vee B=0,
\]
where $\mathrm d_A^\vee\colon\Omega^k(M,\mathfrak g^\vee)\to\Omega^{k+1}(M,\mathfrak g^\vee)$ is the covariant derivative for the coadjoint representation, defined by $\langle \mathrm d_A^\vee\beta,\xi\rangle = \mathrm d\langle\beta,\xi\rangle - \langle\beta,\mathrm d_A\xi\rangle$ for any $\mathfrak g$-valued section $\xi$. The classical solutions are thus flat $G$-connections paired with covariantly closed $B$-fields. The action admits two towers of gauge symmetry. Ordinary gauge transformations, parametrized by $\epsilon\in\Omega^0(M,\mathfrak g)$, act by
\[
  \delta_\epsilon A = \mathrm d_A\epsilon,\qquad \delta_\epsilon B = \mathrm{ad}^*_\epsilon B,
\]
where $\mathrm{ad}^*_\epsilon$ denotes the coadjoint $\mathfrak{g}$-action, $\langle\mathrm{ad}^*_\epsilon\lambda,\xi\rangle = -\langle\lambda,[\epsilon,\xi]\rangle$. The additional BF shift symmetry, parametrized by $\beta\in\Omega^{n-2}(M,\mathfrak g^\vee)$, acts by
\[
  \delta_\beta A = 0,\qquad \delta_\beta B = \mathrm d_A^\vee\beta.
\]
This large gauge symmetry group eliminates all local degrees of freedom, making the theory purely topological.

From the derived-geometric perspective, the shifted cotangent stack $\Tcot{n}{\mathrm B G}$ encodes exactly this field content: \[M \mapsto \Map(M_\flat, \Tcot{n}{\mathrm B G}) \simeq T^* \Map(M_\flat, \mathrm BG).\] The mapping stack $\Map(M_\flat, \mathrm BG)$ parametrizes flat $G$-connections on $M$ modulo gauge transformations, while the cotangent fiber supplies the $B$-field. The canonical $n$-shifted symplectic form on $\Tcot{n}{\mathrm B G}$ recovers, via transgression to $\Map(M_\flat,\Tcot{n}{\mathrm B G})$, a $(-1)$-shifted symplectic structure whose underlying BV bracket is that of the BF action functional.

\subsection{Hamiltonian Actions and Symplectic Reduction}

We recall the classical mechanism by which symmetries are incorporated into symplectic geometry; the treatment here follows \cite{Souriau1969, MarsdenWeinstein1974, Meyer1973} for the foundational material and \cite{GuilleminSternberg1984, MarsdenRatiu1999} for later developments.

Let \((M,\omega)\) be a symplectic manifold and let a Lie group \(G\) act on \(M\) by symplectomorphisms. For each \(\xi\in\mathfrak{g}\), denote by \(\xi_M\) the fundamental vector field on \(M\) induced by \(\xi\). Since the action is by symplectomorphisms, the Lie derivative \(\mathcal{L}_{\xi_M}\omega = 0\), so Cartan's formula gives \(\mathrm{d}(\iota_{\xi_M}\omega)=0\): each \(\iota_{\xi_M}\omega\) is a closed \(1\)-form. The action is called \emph{Hamiltonian} if these closed forms are globally exact in a manner compatible with the Lie bracket, i.e. there exists a \(G\)-equivariant smooth map
\[
  \mu\colon M\longrightarrow \mathfrak{g}^\vee,
\]
called the \emph{moment map}, such that
\[
  \mathrm{d}\langle\mu,\xi\rangle = \iota_{\xi_M}\omega,
  \qquad \forall \xi\in\mathfrak{g}.
\]
Here \(\langle\mu,\xi\rangle := \mu(\,\cdot\,)(\xi)\colon M\to\mathbb{R}\) is the Hamiltonian function for the infinitesimal action of \(\xi\), and equivariance means \(\mu\circ g = \mathrm{Ad}^*_g\circ\mu\) for all \(g\in G\), where \(\mathrm{Ad}^*\) is the coadjoint action on \(\mathfrak{g}^\vee\). This equivariance is automatic when \(G\) is connected and the action is exact, but in general it is an additional datum introduced by Souriau \cite{Souriau1969}. The moment map simultaneously encodes the conserved quantities associated with the symmetry (via Noether's theorem) and the constraint data used in reduction.

The Marsden--Weinstein--Meyer reduction theorem \cite{MarsdenWeinstein1974, Meyer1973} states that a Hamiltonian symmetry should be removed not by the naive orbit quotient \(M/G\), which need not be symplectic, but by first restricting to a level set of \(\mu\) and then quotienting. Concretely, suppose \(0\in\mathfrak{g}^\vee\) is a regular value of \(\mu\) and \(G\) acts freely and properly on \(\mu^{-1}(0)\). Then the \emph{symplectic reduction}
\[
  M /\!/ G \;:=\; \mu^{-1}(0)/G
\]
is a smooth manifold that carries a unique symplectic form \(\omega_{\mathrm{red}}\) determined by
\[
  \pi^*\omega_{\mathrm{red}} = \iota^*\omega,
\]
where \(\iota\colon\mu^{-1}(0)\hookrightarrow M\) is the inclusion and \(\pi\colon\mu^{-1}(0)\to\mu^{-1}(0)/G\) is the quotient. The reason here is that $\ker\mathrm{d}\mu_p = (\mathfrak{g}\cdot p)^{\omega}$ along the level set, where $(-)^\omega$ denotes the symplectic orthogonal complement with respect to $\omega$, so the symplectic orthogonal of the gauge orbit is exactly the tangent space to \(\mu^{-1}(0)\), and the restriction of \(\omega\) descends. More generally, reducing at a coadjoint orbit \(\mathcal{O}\subset\mathfrak{g}^\vee\) gives
\[
  \mu^{-1}(\mathcal{O})/G
  \;\simeq\;
  \mu^{-1}(\lambda)/G_\lambda,
  \qquad \forall \lambda\in\mathcal{O},
\]
where \(G_\lambda\) is the stabilizer of \(\lambda\), carrying the symplectic form obtained by
twisting by the Kirillov--Kostant--Souriau form on \(\mathcal{O}\).

This construction is the classical prototype for Hamiltonian reduction: a symplectic manifold with symmetry produces a new symplectic manifold of constrained solutions modulo gauge. In the derived and shifted setting of this paper, the same idea is realized as a derived intersection
\[
  \qstack{X}{G}\times_{\Tcot{n}{\mathrm{B}G}} \mathfrak{g}^\vee[n-1]
\]
with a conormal Lagrangian; the classical formula \(\mu^{-1}(0)/G\) is recovered as the underived transverse case of this intersection (cf.\ Conjecture~\ref{conj:main} and the discussion in §\ref{sec:classical_continuous_SymTFT}).

\section{General Philosophy}

In this section we summarize the general philosophy of the classical continuous symmetry TFT, and explain the reasons for the various ingredients in the main conjecture.

\subsection{TFTs and Cobordism Hypothesis}

An $n$-dimensional \emph{topological field theory}, in the sense of Atiyah \cite{Atiyah88}, is a symmetric monoidal functor\footnote{In fact, Atiyah also requires the TFT to send orientation reversal of bordisms to the complex conjugation of the Hilbert spaces. In other word, it is equipped with an additional $\mathbb{Z}/2$-equivariant functor structure. This is related to the idea of \emph{unitarity} and \emph{reflection positivity}, see discussion in \cite{FreedHopkins21}.} from the oriented bordisms to Hilbert spaces:
\[\mathcal{T} \colon \operatorname{Bord}^{\mathrm{SO}}_{\langle n,n-1 \rangle} \to \operatorname{Hilb},\] such that the juxtaposition and gluing of bordisms correspond to the tensor product and composition of linear maps.

Baez and Dolan proposed extending this framework to fully extended TFTs, in which bordisms can be cut along corners of arbitrary codimension and the target is a symmetric monoidal $n$-category \cite{BaezDolan95}. They conjectured the \emph{Cobordism Hypothesis}: fully extended framed TFTs with values in a symmetric monoidal $n$-category $\mathcal{C}$ are classified by the fully dualizable objects of $\mathcal{C}$, by evaluation at a point. Lurie gave a rigorous $(\infty,n)$-categorical formulation of fully extended TFTs and provided a sketch proof of the Cobordism Hypothesis \cite{Lurie09}.

\begin{theorem}[Lurie]
  There is an equivalence of $\infty$-groupoids between fully extended $n$-dimensional framed TFTs valued in a symmetric monoidal $(\infty,n)$-category $\mathcal{C}$ and fully dualizable objects of $\mathcal{C}$:
  \begin{equation}
  \operatorname{Fun}_{\mathbb{E}_\infty}(\operatorname{Bord}^{\mathrm{fr}}_{n}, \mathcal{C}) \simeq \operatorname{Obj}^{\mathrm{fd}}(\mathcal{C}).
  \end{equation}
\end{theorem}

\subsection{TFTs with Symmetries}

Given a group $G$ and a representation $\zeta \colon G \to \mathrm{O}(n)$, a fully extended $n$-dimensional $(G,\zeta)$-symmetric TFT valued in $\mathcal{C}$ is a symmetric monoidal functor
\[\mathcal{T} \colon \operatorname{Bord}^{(G,\zeta)}_{n} \to \mathcal{C}.\]
Here, $\operatorname{Bord}^{(G,\zeta)}_{n}$ is the $(\infty,n)$-category of fully extended $n$-bordisms equipped with \emph{tangential $(G,\zeta)$-structures}. 

A tangential $(G,\zeta)$-structure on a $k$-manifold $M$ (for $0 \leq k \leq n$) consists of a principal $G$-bundle $P \to M$, together with a vector bundle isomorphism between $P \times_G \zeta$ and $TM \oplus (\mathbb{R}^{n-k} \times M)$.

\begin{theorem}[Lurie]
  Fully extended $n$-dimensional $(G,\zeta)$-symmetric TFTs valued in $\mathcal{C}$ are classified by the homotopy fixed points on the space of fully dualizable objects of $\mathcal{C}$:
  \begin{equation}
    \operatorname{Fun}_{\mathbb{E}_\infty}(\operatorname{Bord}^{(G,\zeta)}_{n}, \mathcal{C}) \simeq \operatorname{Map}_{G}(\pt, \operatorname{Obj}^{\mathrm{fd}}(\mathcal{C})).
  \end{equation}
\end{theorem}

\subsection{Gauging TFTs}
Gauging is the reversible process of producing a new TFT from a given TFT with symmetry by summing over the background gauge fields. In the fully extended setting, gauging is implemented by the following procedure due to Lurie \cite{Lurie09}.

Take a group $G$ and a representation $\zeta \colon G \to \mathrm{O}(n)$. There is a canonical functor $\operatorname{Bord}^{(G,\zeta)}_{n} \to \operatorname{Bord}^{\mathrm{O}}_{n}$ simply forgetting the tangential $(G,\zeta)$-structures. Then the universal approximation to the extension problem
\begin{equation}
\begin{tikzcd}
  \operatorname{Bord}^{(G,\zeta)}_{n} 
    \arrow[r,"\mathcal{T}"] 
    \arrow[d]
  & \mathcal{C}
  \\ \operatorname{Bord}^{\mathrm{O}}_{n}
\end{tikzcd}
\end{equation}
is given by the left Kan extension symmetric monoidal functor \[\Qstack{\mathcal{T}}{G} \colon \operatorname{Bord}^{\mathrm{O}}_{n} \xrightarrow{\qstack{\mathcal{T}}{G} } \operatorname{Corr}_n(\operatorname{Loc}_{\mathrm{B} G}(\mathcal{C})) \xrightarrow{\operatorname{FM}} \operatorname{Loc}_{\mathrm{B} G}(\mathcal{C}) \xrightarrow{\operatorname{forget}}\mathcal{C},\] where 
\begin{enumerate}[nosep,label=(\roman*)]
  \item $\operatorname{Loc}_{\mathrm{B} G}(\mathcal{C}) := \operatorname{Fun}(\mathrm{B} G, \mathcal{C})$ is the symmetric monoidal $(\infty,n)$-category of $\mathcal{C}$-valued local systems on the classifying stack $\mathrm{B} G$, 
  
  \item $\operatorname{Corr}_n(-)$ is the functorial process which sends $(\infty,n)$-categories to their $n$-fold correspondence categories, 
  
  \item $\operatorname{FM}$ is the higher-categorical version of the \emph{Fourier--Mukai transform} given by pullback-pushforward along the correspondences.
\end{enumerate}

\subsection{Finite SymTFT}
On the other hand, the program of gauging symmetries via topological defects in higher-dimensional topological field theory has seen remarkable progress over the last decade, particularly for finite symmetries and their generalizations to higher fusion categories.

For simplicity, let us consider 2-dimensional topological field theories with finite group symmetry $G$ and a 't Hooft anomaly $\omega\in\operatorname{H}^3(G;\mathbb{C}^\times)$. The target 2-category is the Morita 2-category of algebras, bimodules and intertwiners. It is well-known\footnote{See \cite{SchommerPries09} for more details.} that framed 2-dimensional TFTs in this target are classified by separable algebras, or equivalently, enumerated by a positive integer $n$ corresponding to the number of simple objects in the category of boundary conditions. The data of an anomalous $G$-action on such a TFT via topological line defects is encoded by a monoidal functor 
\begin{equation} 
  F \colon \mathrm{Vec}_G^\omega \to \operatorname{End}(\mathrm{Vec}^{\oplus n}),
\end{equation} 
where $\mathrm{Vec}_G^\omega$ is the fusion category of $G$-graded vector spaces with associator twisted by $\omega$. 

The symmetry TFT in this case is the 3-dimensional Dijkgraaf--Witten theory with gauge group $G$ and twist $\omega$. Wilson lines in this bulk TFT form a modular tensor category, namely the Drinfeld center $\mathcal{Z}(\mathrm{Vec}_G^\omega)$. Topological boundary conditions are classified by Lagrangian algebras in $\mathcal{Z}(\mathrm{Vec}_G^\omega)$, which are in bijection with conjugacy classes of pairs $(H,\psi)$, where $H \leq G$ is a subgroup and $\psi\in \operatorname{C}^2(H;\mathbb{C}^\times)$ is a 2-cochain satisfying $\mathrm{d}\psi = \omega|_H$. Topological line defects on the boundary classified by $(H,\psi)$ form a fusion category $\mathrm{Vec}_{H \backslash G/H}^{\widetilde{\omega}}$, given by the category of $\mathbb{C}[H,\psi]$-bimodules in $\mathrm{Vec}_G^\omega$.

In particular, the Dirichlet boundary condition corresponds to $H = *$ and $\psi = 1$, while the Neumann boundary condition only exists when $\omega$ is trivial, and corresponds to $H = G$ and $\psi = 1$. 

In this paradigm, the physical boundary condition is determined by the collection of topological line defects, which forms the fusion category $\operatorname{End}_{\mathrm{Vec}_G^\omega}(\mathrm{Vec}^{\oplus n})$. The topological line defects in the original theory are recovered by closing the sandwich with the Dirichlet boundary: 
\begin{equation}
  \operatorname{End}(\mathrm{Vec}^{\oplus n}) \simeq \operatorname{End}_{\mathrm{Vec}_G^\omega}(\mathrm{Vec}^{\oplus n}) \boxtimes_{\mathcal{Z}(\mathrm{Vec}_G^\omega)} \mathrm{Vec}_G^\omega. 
\end{equation}

Therefore, to gauge the $G$-symmetry to a subgroup $H$, one simply replaces the topological boundary condition on the symmetry side by the one corresponding to $H$. The 2-cochain $\psi$ is an extra redundancy in this gauging process, which is related to stacking a 2-dimensional $H$-SPT phase when the anomaly $\omega|_H$ is trivial. The resulting gauged theory has topological line defects given by 
\begin{equation}
  \operatorname{Bimod}_{\operatorname{End}(\mathrm{Vec}^{\oplus n})}(\mathbb{C}[H,\psi]) \simeq \operatorname{End}_{\mathrm{Vec}_G^\omega}(\mathrm{Vec}^{\oplus n}) \boxtimes_{\mathcal{Z}(\mathrm{Vec}_G^\omega)} \mathrm{Vec}_{H \backslash G/H}^{\widetilde{\omega}},
\end{equation} 
which is Morita dual to the category of line defects in the original theory, while topological lines living on the topological domain wall form the invertible bimodule category 
\begin{equation}
  \operatorname{Mod}_{\operatorname{End}(\mathrm{Vec}^{\oplus n})}(\mathbb{C}[H,\psi]) \simeq \operatorname{End}_{\mathrm{Vec}_G^\omega}(\mathrm{Vec}^{\oplus n}) \boxtimes_{\mathcal{Z}(\mathrm{Vec}_G^\omega)} \mathrm{Vec}_{G/H}. 
\end{equation}

\subsection{Connection between the Two Perspectives of Gauging}
Recall that a fully extended 2-dimensional $G$-symmetric TFT $\mathcal{T}$ is a symmetric monoidal functor $\operatorname{Bord}^{(G,\zeta)}_{2} \to \mathrm{2Vec}$ where $\zeta$ is trivial. If we carefully unpack the definitions\footnote{See \cite{Turaev10} for details.}, we find that $\mathcal{T}$ is determined by a $G$-graded separable algebra $A = \bigoplus_{g \in G} A_g$, i.e., the multiplication is restricted to $A_g \otimes A_h \to A_{gh}$ for all $g,h \in G$ and the unit lies in $A_e$. These data are exactly the image of the group algebra $\mathbb{C}[G]$ under the monoidal functor $F \colon \mathrm{Vec}_G \to \operatorname{End}(\mathrm{Vec}^{\oplus n})$. Thus, gauging the $G$-symmetry to $(H,\psi)$ corresponds to identifying a subalgebra of $A$ with the image of $\mathbb{C}[H,\psi]$ under $F$. 

For any separable algebra $A_0$ with a $G$-action $\rho \colon G \to \operatorname{Aut}(A_0)$, we can consider the induced $G$-action on the corresponding TFT via 
\begin{equation}
  G \xrightarrow{\rho} \operatorname{Aut}(A_0) \xrightarrow{[-]} \operatorname{Bimod}(A_0). 
\end{equation} 
Algebraically, this is equivalent to a $G$-graded extension $A = \bigoplus_{g \in G} A_g$ where $A_0 = A_e$, in the sense of Etingof--Nikshych--Ostrik \cite{ENO10}. Recall that the physical boundary corresponds to a Lagrangian algebra in $\mathcal{Z}(\mathrm{Vec}_G)$, or more concretely, a \emph{$G$-crossed commutative algebra}. For any $G$-graded algebra $A$, we can construct a $G$-crossed commutative algebra by taking its \emph{full center} in the sense of Davydov \cite{Davydov10}.

Note that the induced $G$-action $[\rho]$ on the 2-dimensional TFT admits a trivialization if and only if the $G$-action $\rho$ on the algebra $A_0$ is \emph{inner}, i.e., the action $\rho$ can be lifted to a homomorphism $\widetilde{\rho} \colon G \to A_0^\times$ to the group of units of $A_0$. Gauging such a trivial $G$-action, in the SymTFT picture, corresponds to decoupling the physical boundary as a Neumann boundary stacked with a stand-alone 2-dimensional TFT.

When the 't Hooft anomaly $\omega$ is nontrivial, we cannot simply consider fully extended 2-dimensional $G$-symmetric TFTs. To accommodate the anomaly, we need to consider a 3-dimensional $G$-SPT, namely an \emph{invertible TFT} $\mathcal{I}^\omega$ with $G$-symmetry, and then consider fully extended 2-dimensional \emph{relative TFTs} with respect to $\mathcal{I}^\omega$ in the sense of Freed--Teleman \cite{FreedTeleman14}. The SymTFT, which is the 3-dimensional Dijkgraaf--Witten theory with gauge group $G$ and twist $\omega$, is exactly the TFT obtained by gauging the $G$-symmetry of $\mathcal{I}^\omega$. The physical boundary condition is then obtained via gauging the $G$-symmetry of the original relative TFT.

\section{Classical Continuous SymTFT}\label{sec:classical_continuous_SymTFT}
Now we are ready to explain the general picture of the classical continuous symmetry TFT. Let $G$ be an algebraic group (in the derived algebraic geometric setting) or a Lie group (in the setting of derived differential stacks). We consider $G$ as the \emph{internal symmetry}, which means the representation $\zeta \colon G \to \mathrm{SO}(n)$ is trivial. 

\subsection{Functorial Field Theory Framework} A fully extended classical $n$-dimensional $G$-symmetric TFT takes values in the symmetric monoidal $(\infty,n)$-category $\mathrm{Lag}_n$ of shifted Lagrangians $n$-fold correspondences from Calaque \cite{Calaque15}, namely, a symmetric monoidal functor:
\begin{equation}
  \mathcal{T} \colon \operatorname{Bord}^{(G,\zeta)}_{n} \to \mathrm{Lag}_n. 
\end{equation}

By the Cobordism Hypothesis, $\mathcal{T}$ is determined by its value on a point, which is an $(n-1)$-shifted symplectic stack $(X,\omega)$ equipped with a $G$-action via topological defects:
\begin{equation}
  [\rho] \colon G \to \operatorname{Aut}_{\mathrm{Lag}_{n}}(X,\omega).
\end{equation}
Here, we assume this $G$-action is induced from a $G$-action $\rho$ on $X$ by symplectomorphisms, namely, it is given by the composition
\begin{equation}
  G \xrightarrow{\rho} \operatorname{Aut}_{\mathrm{Symp}_{n-1}}(X,\omega) \xrightarrow{[-]} \operatorname{Aut}_{\mathrm{Lag}_{n}}(X,\omega).
\end{equation}
Note that by the Grothendieck construction, the $G$-action on the stack $X$ is encoded by a map from the quotient stack $\qstack{X}{G}$ to the classifying stack $\mathrm{B} G$. Consider the shifted conormal stack
\begin{equation}
  N^*[n](\qstack{X}{G} \to \mathrm{B} G) \to \Tcot{n}{\mathrm{B} G}.
\end{equation}
We claim that this shifted conormal stack is precisely the physical boundary of the SymTFT; that is, it carries a canonical structure of a shifted Lagrangian in $\Tcot{n}{\mathrm{B} G}$.

To see why, recall from the general theory of shifted symplectic structures \cite{PTVV13,Calaque15} that for any map of derived stacks $f \colon Y \to Z$, the shifted conormal stack $N^*[k](Y \to Z)$ admits a Lagrangian structure relative to $\Tcot{k}{Z}$ whenever $Z$ carries a $k$-shifted symplectic structure. Since the bulk target $\Tcot{n}{\mathrm{B}G}$ is canonically $n$-shifted symplectic via the cotangent pairing, the shifted conormal stack $N^*[n](\qstack{X}{G} \to \mathrm{B}G)$ acquires an $n$-shifted Lagrangian structure mapping to $\Tcot{n}{\mathrm{B}G}$. This is the physical boundary.

It remains to verify that this boundary encodes the correct physical data.
\begin{enumerate}
  \item The classifying map for the bare $G$-action $\qstack{X}{G} \to \mathrm{B}G$ induces, by functoriality of shifted cotangents, a canonical \emph{cotangent correspondence}
  \[
    \Tcot{n}{\qstack{X}{G}} \longleftarrow \Tcot{n}{\mathrm{B}G} \times_{\mathrm{B}G} \qstack{X}{G} \longrightarrow \Tcot{n}{\mathrm{B}G}.
  \]
  This is a Lagrangian correspondence using only the bare action, with no symplectic data on $X$ required.

  \item The $G$-invariance of the $(n-1)$-shifted symplectic form $\omega$ on $X$ determines a section of $\Tcot{n}{\qstack{X}{G}}$, namely the graph of $\omega$ descended to the quotient $\qstack{X}{G}$. Composing this section with the cotangent correspondence of step~(1) collapses the $\Tcot{n}{\qstack{X}{G}}$ leg to a point and yields an honest Lagrangian map into $\Tcot{n}{\mathrm{B}G}$, which is exactly the shifted conormal stack $N^*[n](\qstack{X}{G} \to \mathrm{B}G)$.

  \item  Closing the SymTFT sandwich verifies the identification. The sandwich is formed by taking the derived fiber product of the physical boundary with each topological boundary of $\Tcot{n}{\mathrm{B}G}$;

  \item Closing against the \emph{Dirichlet boundary} (the cotangent fiber $\mathfrak{g}^\vee[n-1] \hookrightarrow \Tcot{n}{\mathrm{B}G}$) recovers the ungauged theory: the derived fiber product is $X$ with its $(n-1)$-shifted symplectic structure and residual $G$-action;
  
  \item Closing against the \emph{Neumann boundary} (the zero section $\mathrm{B}G \hookrightarrow \Tcot{n}{\mathrm{B}G}$) recovers the gauged theory: the derived fiber product is the $(n-1)$-shifted symplectic reduction $\Qstack{X}{G}$.
\end{enumerate}

\subsection{Hamiltonian Reduction}
If the $G$-action $[\rho]$ on $X$ has a trivialization, then the action $\rho$ lifts to a shifted \emph{Hamiltonian} action:
\begin{equation}
  \widetilde{\rho} \colon G \to \operatorname{Aut}_{\mathrm{Ham}_{n-1}}(X,\omega),
\end{equation}
where the target is \emph{defined} to be the kernel of the homomorphism 
\begin{equation}
  [-] \colon \operatorname{Aut}_{\mathrm{Symp}_{n-1}}(X,\omega) \to \operatorname{Aut}_{\mathrm{Lag}_{n}}(X,\omega).
\end{equation}

To justify this definition, we will extract the shifted version of the moment map from the trivialization data. 

The shifted moment map is exactly the data measuring whether the physical boundary $N^*[n](\qstack{X}{G} \to \mathrm{B}G)$ can be \emph{descended} to a second Lagrangian of $\Tcot{n}{\mathrm{B}G}$ whose underlying stack is the bare quotient $\qstack{X}{G}$ itself. Concretely, write $p \colon \qstack{X}{G} \to \mathrm{B}G$ for the classifying map and consider the cotangent projection
\begin{equation}
  \pi \colon \Tcot{n}{\mathrm{B}G} \to \mathrm{B}G.
\end{equation}
A lift of $p$ through $\pi$ is by definition a section of the pulled-back cotangent complex $p^*\mathbb{L}_{\mathrm{B}G}[n] \simeq p^*\mathfrak{g}^\vee[n-1]$ over $\qstack{X}{G}$, i.e. a $G$-equivariant map
\begin{equation}
  \mu \colon X \to \mathfrak{g}^\vee[n-1],
\end{equation}
which is precisely the \emph{$(n-1)$-shifted moment map}. The zero section $\mathrm{B}G \hookrightarrow \Tcot{n}{\mathrm{B}G}$ furnishes the lift $\mu = 0$, but its isotropy is the only structure that comes for free: a lift is a genuine \emph{Lagrangian map} $\qstack{X}{G} \to \Tcot{n}{\mathrm{B}G}$ if and only if $\mu$ is nondegenerate, that is, if and only if the moment map equation 
\begin{equation}
  \iota_{\rho(\xi)}\omega = d\langle \mu, \xi\rangle
\end{equation}
and the bracket condition
\begin{equation}
  \langle \mu, [\xi,\eta]\rangle = \{\mu_\xi, \mu_\eta\}
\end{equation} both hold. Together, they are the Hamiltonian condition. The moment map is exactly the trivialization datum that descends the physical boundary from the conormal stack to the quotient stack $\qstack{X}{G}$.

If the bracket condition fails, the failure is measured by the curvature cocycle $c(\xi,\eta) = \mu_{[\xi,\eta]} - \{\mu_\xi,\mu_\eta\}$. Its natural home is Chevalley--Eilenberg cohomology of $\mathfrak{g}$ with coefficients in the \emph{shifted center} of the shifted Poisson algebra of functions on $X$,
\begin{equation}
  \mathfrak{Z}^\bullet := \mathrm{HP}^\bullet(\mathcal{O}_X),
\end{equation}
the complex of shifted Casimirs (the derived centre of the bracket $\{-,-\}$). We emphasise that, unlike the ordinary Poisson case where the Casimirs form a single vector space in degree $0$, here $\mathfrak{Z}^\bullet$ is a genuine complex with components in several internal degrees, reflecting the $(n-1)$-shift of $\omega$. The obstruction therefore lives in the \emph{total} complex $C^\bullet\bigl(\mathfrak{g},\, \mathfrak{Z}^\bullet\bigr)$, whose total degree is the sum of the Chevalley--Eilenberg degree and the internal degree of $\mathfrak{Z}^\bullet$; the curvature is the \emph{total-degree-$2$} class
\begin{equation}
  [c]\in \mathrm{H}^2\bigl(C^\bullet(\mathfrak{g},\, \mathfrak{Z}^\bullet)\bigr).
\end{equation}
In the unshifted limit $\mathfrak{Z}^\bullet$ collapses to the Casimir functions in degree $0$, the total degree reduces to the Chevalley--Eilenberg degree, and one recovers the classical class $[c]\in \mathrm{H}^2(\mathfrak{g}, \mathcal{Z}(X))$; in the shifted setting the same total-degree-$2$ class may spread its components across several Chevalley--Eilenberg/internal bidegrees.

\subsection{Symplectic Reduction}
When $[c]=0$ the descent is clean: the Dirichlet sandwich $\qstack{X}{G} \times^h_{\Tcot{n}{\mathrm{B}G}} \mathfrak{g}^\vee[n-1]$ recovers the stack $X$, while the Neumann sandwich $\qstack{X}{G} \times^h_{\mathrm{B}G} \mathrm{B}G$ recovers the symplectic reduction $\Qstack{X}{G} \simeq \mu^{-1}(0)/G$. 

When $[c]\neq 0$, the obstruction is resolved one total degree at a time: the curvature $c = c_2$ heads a tower of higher classes $[c_2], [c_3], \dots$, where $[c_k]$ has total degree $k$ in $C^\bullet(\mathfrak{g}, \mathfrak{Z}^\bullet)$ and is absorbed by an antifield of ghost degree $-(k-1)$ (so $c_2$ by an \emph{antifield}, $c_3 = [\partial c]$ by an \emph{antighost}, and so on). These data assemble into an $L_\infty$-algebra structure on $\mathfrak{g}[1]\oplus \mathfrak{Z}^\bullet$, with $\ell_1$ the Chevalley--Eilenberg differential, $\ell_2$ the action bracket, and $\ell_k$ ($k\geq 3$) the higher bracket carrying $c_k$.

Following the AKSZ description \cite{AKSZ97} of this $L_\infty$-algebra, the obstruction tower is packaged into a single object. Let $\mathcal{F}$ denote the derived field space whose functions are the BRST/ghost complex
\begin{equation}
  \mathcal{O}_X \otimes \operatorname{Sym}(\mathfrak{g}[-1] \oplus \mathfrak{g}^\vee[1]),
\end{equation}
and form the shifted cotangent $\Tcot{-1}{\mathcal{F}}$, the \emph{field--antifield} space, equipped with its canonical $(-1)$-shifted Poisson bracket (the antibracket) $\{-,-\}$. The Maurer--Cartan element is then a distinguished degree-$0$ function
\begin{equation}
  S \;=\; \underbrace{d_{\mathrm{CE}}}_{\ell_1} \;+\; \underbrace{\langle \mu,\,\xi\rangle}_{\ell_2} \;+\; \sum_{k\geq 3} \underbrace{\tfrac{1}{k!}\,\langle c_k,\, \xi^{\otimes k}\rangle}_{\ell_k}\ \in\ \mathcal{O}\bigl(\Tcot{-1}{\mathcal{F}}\bigr),
\end{equation}
where $\xi$ ranges over the ghost generators $\mathfrak{g}[1]$, the $k$-th term pairs the obstruction cochain $c_k$ against $k$ ghosts, and the antifields enter through the terms required to make $S$ have ghost degree $0$. By construction $S$ is the AKSZ Hamiltonian, and the statement that it is a Maurer--Cartan element of the $L_\infty$-algebra is exactly the BV \emph{master equation}
\begin{equation}
  \{S,S\}=0,
\end{equation}
the nilpotency $Q^2 = 0$ of the BV differential $Q = \{S,-\}$.

In this case, the Neumann sandwich is the derived reduction $\Qstack{X}{G}$ carrying the full BV/$L_\infty$ structure. There is a spectral sequence of the antifield filtration, which collapses at the $E_2$ page exactly when $[c]=0$, is the algebraic shadow of this derived intersection. The Hamiltonian condition is therefore never needed to \emph{set up} the sandwich, which exists for any symplectic action; it controls only whether the physical boundary descends to an honest shifted moment map $\mu$ on $\qstack{X}{G}$, or remains an irreducible conormal correspondence carrying its $L_\infty$ tail.

\subsection{Lagrangian Reduction}
In general, a $G$-symmetry on a fully extended $n$-dimensional oriented TFT, determined by an $(n-1)$-shifted symplectic stack $(X,\omega)$, is not necessarily induced by a $G$-action on $X$ by symplectomorphisms. 

Nevertheless, the symmetry TFT paradigm should still apply, and this suggests that there is an even more general notion of reduction via invertible Lagrangian correspondences, which we call \emph{Lagrangian reduction}. The main goal is to extract the physical boundary of the SymTFT as a shifted Lagrangian in $\Tcot{n}{\mathrm{B}G}$ directly from a homomorphism
\begin{equation}
  \Gamma \colon G \to \operatorname{Aut}_{\mathrm{Lag}_{n}}(X,\omega).
\end{equation}

Since $G$ is a group object internal to the category of derived stacks (i.e., $G$ is an algebraic group if we work in derived algebraic geometry, or a Lie group if we work in derived differential geometry), we can construct a $G$-weighted colimit of $\Gamma$ by \emph{tensoring} over $G$: \[\int_{g \in G} \Gamma(g).\]
Note that the source and target of the Lagrangian $\Gamma(g)$ are both $X$ for each $g$, so the colimit is equipped with two canonical maps to $X$. Meanwhile, there is a canonical map from $X$ to the colimit, given by the unit of the $G$-action.

Then the group structure on $G$ enables us to extend these maps to a simplicial object in the category of derived stacks:
\begin{equation}\label{eqn:Lagrangian_resolution}
  \begin{tikzcd}
    {X}
      \arrow[r]
    & {\int_{g \in G} \Gamma(g)}
      \arrow[l,shift left=1ex]
      \arrow[l,shift right=1ex]
      \arrow[r,shift left=1ex]
      \arrow[r,shift right=1ex]
    & {\int_{g,h \in G} \Gamma(g) \times \Gamma(h) \times \Gamma(gh)}
      \arrow[l,shift left=2ex]
      \arrow[l,shift right=2ex]
      \arrow[l]
      \arrow[r,shift left=2ex]
      \arrow[r,shift right=2ex]
      \arrow[r]
    & {\cdots}
      \arrow[l,shift left=3ex]
      \arrow[l,shift right=3ex]
      \arrow[l,shift left=1ex]
      \arrow[l,shift right=1ex]
  \end{tikzcd}
\end{equation}
Then the \emph{geometric realization}, i.e., the colimit of this simplicial object, is the geometric shape of the desired physical boundary of the SymTFT, which we also denote by $\qstack{X}{G}$. In particular, when $\Gamma \simeq [\rho]$ is induced from a $G$-action on $X$ by symplectomorphisms, the above construction recovers the action groupoid
\begin{equation}\label{eqn:action_groupoid}
  \begin{tikzcd}
    {X}
      \arrow[r]
    & {X \times G}
      \arrow[l,shift left=1ex]
      \arrow[l,shift right=1ex]
      \arrow[r,shift left=1ex]
      \arrow[r,shift right=1ex]
    & {X \times G \times G}
      \arrow[l,shift left=2ex]
      \arrow[l,shift right=2ex]
      \arrow[l]
      \arrow[r,shift left=2ex]
      \arrow[r,shift right=2ex]
      \arrow[r]
    & {\cdots}
      \arrow[l,shift left=3ex]
      \arrow[l,shift right=3ex]
      \arrow[l,shift left=1ex]
      \arrow[l,shift right=1ex]
  \end{tikzcd}
\end{equation}
so we recover the previous construction of the symplectic reduction as a special case of this more general Lagrangian reduction.

Notice that the simplicial object \eqref{eqn:Lagrangian_resolution} still admits a simplicial map to $\mathrm{B} G$, realized as the trivial action groupoid \eqref{eqn:action_groupoid}. To realize the physical boundary as a Lagrangian in $\Tcot{n}{\mathrm{B}G}$, we apply the cotangent stack functor $\Tcot{n}(-)$ to \eqref{eqn:Lagrangian_resolution} to obtain a Lagrangian correspondence, and then take the derived intersection with the zero section of the shifted cotangent stack, which generalizes the shifted conormal stack construction in the symplectic case. 

\subsection{Twisting the SymTFT}
Let us now briefly discuss the case when the $G$-action is \emph{anomalous}. By analogy with the finite case, we should consider an invertible one-dimensional higher $G$-TFT $\mathcal{I}^\omega$ together with a $G$-symmetric relative TFT $\mathcal{T}$ with respect to $\mathcal{I}^\omega$. In the functorial field theory setting this requires a distinction between two targets.

The target $\Lag_n$ used above is the \emph{classical} target. Its objects are shifted symplectic stacks and its morphisms are shifted Lagrangian correspondences. It remembers closed shifted forms and isotropic homotopies, but it does not remember integral lifts (\emph{Maslov index}) or prequantum phases. 

Recall that an \emph{$n$-gerbe} on a stack $Y$ may be viewed as a principal $\mathrm{B}^n\mathbb{G}_m$-bundle over $Y$, or equivalently as a class in $\operatorname{H}^{n+1}(Y;\mathbb{G}_m)$, up to the usual indexing conventions \cite{ToenVezzosi08,NikolausSchreiberStevenson15}. Its differential refinement is described by higher differential cohomology, equivalently by higher gerbes with connection \cite{Schreiber13}.

There is a \emph{prequantum} refinement of the shifted Lagrangian correspondence category
\begin{equation}
  \PreqLag_n \to \Lag_n
\end{equation}
whose objects are shifted symplectic stacks equipped with a prequantum higher gerbe with connection whose curvature is the shifted symplectic form, and whose Lagrangian correspondences are equipped with compatible trivializations of the pulled-back gerbes \cite{Safronov2023}. This is the shifted analogue of passing from a symplectic manifold to a prequantum line bundle, and from a Lagrangian submanifold to a brane equipped with a flat trivialization of the prequantum line. Thus the usual classical isotropic homotopy is the curvature, or shadow, of a prequantum gerbe trivialization:
\begin{center}
\fbox{$\displaystyle\begin{aligned}
  \text{classical isotropic homotopy}
  &=
  \text{curvature of the prequantum trivialization}.
\end{aligned}$}
\end{center}

A \emph{prequantum AKSZ TFT} is a lift of the classical AKSZ TFT, in the sense of Calaque--Haugseng--Scheimbauer \cite{CHS25}, from $\Lag_n$ to $\PreqLag_n$. The forgetful functor discards the gerbe decorations and recovers the classical field theory. The decorations, however, carry exactly the integral and differential cohomology data in which anomaly phases live.

Fully characterizing $\PreqLag_n$ and its invertible objects is beyond the scope of this article, but there is a canonical family of invertible objects that suffices for the symmetry TFT applications. Tensor product of gerbes makes these objects invertible, and prequantum decoration gives a natural inclusion of phases:
\begin{equation}
  \mathrm{B}^{n+1}\mathbb{G}_m \to \PreqLag_{n+1}.
\end{equation}
Thus an $n$-gerbe $\omega \in \mathrm{H}^{n+1}(\mathrm{B} G;\mathbb{G}_m)$ is not merely an external cohomology label: in the prequantum category it is an invertible $(n+1)$-dimensional prequantum field theory $\mathcal{I}^\omega$ with a $G$-action. A relative $n$-dimensional TFT over $\mathcal{I}^\omega$ models a classical $n$-dimensional TFT with an anomalous $G$-symmetry of type $\omega$.

The symmetry TFT sandwich for such an anomalous theory still has the bulk theory given by the shifted cotangent $\Tcot{n}{\mathrm{B}G}$, but now with a nontrivial prequantum decoration determined by the \emph{pullback} of $\omega$ along the cotangent projection map $\Tcot{n}{\mathrm{B}G} \to \mathrm{B}G$. However, notice that topological boundaries of the prequantum bulk are decorated shifted Lagrangians with a stronger notion of trivialization, so the topological boundaries of the classical bulk are not guaranteed to be lifted to the prequantum world. In particular, the Neumann boundary $\mathrm{B}G \hookrightarrow \Tcot{n}{\mathrm{B}G}$ is not expected to admit a prequantum lift, so after closing the sandwich, the gauged theory $\Qstack{X}{G}$ for any $(n-1)$-shifted symplectic stack $(X,\omega)$ with an anomalous $G$-symmetry is not expected to be lifted to a prequantum TFT. 

\subsection{Finite/Quantum and Continuous/Classical Dictionary}

\begin{center}
\footnotesize
\begin{tabular}{p{0.30\linewidth}|p{0.30\linewidth}|p{0.30\linewidth}}
\toprule
\textbf{Concept} & \textbf{Finite/quantum} & \textbf{Continuous/classical} \\
\midrule
TFT target & $n\mathrm{Vec}$, higher Morita category & $\mathrm{Lag}_{n}$, shifted Lagrangians \\
't Hooft anomaly & $G$-SPTs & higher gerbes over $\mathrm{B}G$ \\
Bulk theory & Dijkgraaf--Witten theory &
Classical BF theory  \\
Dirichlet boundary & $\mathrm{Vec}_G$ &
$\mathrm B G\to \Tcot{n}{\mathrm B G}$ \\
Neumann boundary &$\mathrm{Rep}(G)$ &
$\qstack{X}{G}\to \Tcot{n}{\mathrm B G}$ \\
Gauging & modules over $\mathbb{C}[H,\psi]$ &
symplectic reduction \\
trivialization & invertible bimodules & invertible Lagrangians in $X \times \overline{X}$  \\
\bottomrule
\end{tabular}
\end{center}

\section{Examples at \texorpdfstring{\(n=2\)}{n=2}}\label{sec:examples}

Let us focus on the rich case where $n=2$, where the bulk target $\Tcot{2}{\mathrm{B}G}$ is the classical 3-dimensional BF target. In this section, we work at the infinitesimal level, replacing the global stack $\Tcot{2}{\mathrm{B}G}$ by its linearization $\mathrm{B} \, D(\mathfrak{g})$, where $D(\mathfrak{g})$ is the Drinfeld double of the Lie algebra $\mathfrak{g}$. This passage to Lie algebras has two advantages: the relevant structures are entirely classical, and there is a mature classification theory that we can compare directly with our SymTFT dictionary.

\subsection{Manin Pairs, Triples, and the Drinfeld Double}

A \emph{quadratic Lie algebra} is a Lie algebra $\mathfrak{d}$ equipped with a symmetric, nondegenerate, $\mathrm{ad}$-invariant bilinear form $\langle-,-\rangle$. A Lie subalgebra $\mathfrak{l}\subseteq\mathfrak{d}$ is \emph{Lagrangian} if $\mathfrak{l}^\perp=\mathfrak{l}$, or equivalently if $\langle\mathfrak{l},\mathfrak{l}\rangle=0$ and $\dim\mathfrak{l}=\tfrac{1}{2}\dim\mathfrak{d}$.

\begin{definition}
A \emph{Manin pair} $(\mathfrak{d},\mathfrak{g})$ consists of a quadratic Lie algebra $\mathfrak{d}$ and a Lagrangian Lie subalgebra $\mathfrak{g}\subseteq\mathfrak{d}$. A \emph{Manin triple} $(\mathfrak{d},\mathfrak{g},\mathfrak{g}')$ is a Manin pair together with a second Lagrangian subalgebra $\mathfrak{g}'\subseteq\mathfrak{d}$ complementary to $\mathfrak{g}$, i.e.\ $\mathfrak{d}=\mathfrak{g}\oplus\mathfrak{g}'$ as vector spaces.
\end{definition}

Manin triples were introduced by Drinfeld \cite{Drinfeld1993} as the Lie-algebraic encoding of Lie \emph{bialgebras}: a finite-dimensional Lie bialgebra structure on $\mathfrak{g}$ is equivalent to a Manin triple $(\mathfrak{d},\mathfrak{g},\mathfrak{g}')$, uniquely up to isomorphism. In the shifted symplectic language, $\mathrm{B}\mathfrak{d}$ is $2$-shifted symplectic, the inclusion $\mathrm{B}\mathfrak{g}\hookrightarrow\mathrm{B}\mathfrak{d}$ is a 2-shifted Lagrangian, and a Manin triple identifies both $\mathrm{B}\mathfrak{g}$ and $\mathrm{B}\mathfrak{g}'$ as Lagrangians over $\mathrm{B}\mathfrak{d}$.

The canonical example is the \emph{cotangent double}. For any Lie algebra $\mathfrak{g}$, the \emph{Drinfeld double} is the semidirect product
\begin{equation}
  D(\mathfrak{g}) \;:=\; \mathfrak{g}\ltimes\mathfrak{g}^\vee,
\end{equation}
with bracket 
\begin{equation}
  [x+\xi,\,y+\eta]=[x,y]+\mathrm{ad}_x^*\eta-\mathrm{ad}_y^*\xi, \qquad \forall x,y\in\mathfrak{g},\; \forall \xi,\eta\in\mathfrak{g}^\vee,
\end{equation} 
and canonical pairing
\begin{equation}
  \langle x+\xi,\,y+\eta\rangle \;=\; \eta(x)+\xi(y), \qquad \forall x,y\in\mathfrak{g},\; \forall \xi,\eta\in\mathfrak{g}^\vee.
\end{equation}
The subalgebras $\mathfrak{g}$ and $\mathfrak{g}^\vee$ are both Lagrangian, forming the standard Manin triple $(D(\mathfrak{g}),\mathfrak{g},\mathfrak{g}^\vee)$.

At the stack level, this is precisely the infinitesimal model for classical BF theory:
\begin{equation}
  \mathrm{B}\,D(\mathfrak{g}) \;\simeq\; \Tcot{2}{\mathrm{B}\mathfrak{g}}.
\end{equation}
The two Lagrangian summands $\mathfrak{g}^\vee$ and $\mathfrak{g}$ model the linearized Dirichlet and Neumann boundaries in the SymTFT sandwich; Lagrangian subspaces of $D(\mathfrak{g})$ are infinitesimal topological boundary conditions for the 3-dimensional BF theory.

\begin{remark}
This boundary interpretation is not merely formal. Any quadratic Lie algebra $\mathfrak{d}$ induces a Chern--Simons-type action $S_{\mathrm{CS}}(A)=\int_M\tfrac{1}{2}\langle A,\mathrm{d}A\rangle+\tfrac{1}{6}\langle A,[A,A]\rangle$ for closed 3-manifold $M$ with connection $A \in \Omega^1(M;\mathfrak{d})$. 

On a 3-manifold $M$ with boundary, the boundary variation $\frac{1}{2}\int_{\partial M}\langle\delta A,A\rangle$ vanishes whenever $A|_{\partial M}$ takes values in a Lagrangian subspace $\mathfrak{l}\subset\mathfrak{d}$, and gauge invariance of the boundary condition requires $\mathfrak{l}$ to be a Lie subalgebra \cite{Severa16,PullmannSeveraValach19}. In the AKSZ language, a Lagrangian Lie subalgebra $\mathrm{B} \mathfrak{l} \hookrightarrow \mathrm{B}\mathfrak{d}$ is precisely a topological boundary condition for the BF theory.
\end{remark}

\subsection{Topological Boundaries and Their Classification}

The linearized topological boundaries of $\Tcot{2}{\mathrm{B}\mathfrak{g}}$ are classified by Lagrangian Lie subalgebras of $D(\mathfrak{g})$. Following the SymTFT dictionary, each such subalgebra corresponds to an infinitesimal gauging of some subgroup $H\leq G$.

\begin{proposition}[\cite{Courant90,Drinfeld1993,Gualtieri04}]\label{prop:lag-subalg-cotangent}
Every Lagrangian Lie subalgebra of $D(\mathfrak{g})=\mathfrak{g}\ltimes\mathfrak{g}^\vee$ is of the form
\begin{equation}
  \mathfrak{l}_{(\mathfrak{h},\psi)} \;:=\; \bigl\{\, x + \xi \in \mathfrak{h}\oplus\mathfrak{g}^\vee \;\big|\; \xi|_{\mathfrak{h}} = \iota_x\psi \,\bigr\},
\end{equation}
where $\mathfrak{h}\subseteq\mathfrak{g}$ is a Lie subalgebra and $\psi\in \operatorname{Z}^2_{\mathrm{CE}}(\mathfrak{h};\mathbb{C})$ is a Chevalley--Eilenberg $2$-cocycle. Two such subalgebras $\mathfrak{l}_{(\mathfrak{h},\psi)}$ and $\mathfrak{l}_{(\mathfrak{h},\psi')}$ are gauge-equivalent if and only if $\psi-\psi'\in \operatorname{B}^2_{\mathrm{CE}}(\mathfrak{h};\mathbb{C})$. In particular, the equivalence class of the Lagrangian Lie subalgebra is determined by the pair $(\mathfrak{h},[\psi])$ where $[\psi] \in\mathrm{H}^2_{\mathrm{CE}}(\mathfrak{h};\mathbb{C})$.
\end{proposition}

The untwisted case $\psi=0$ gives $\mathfrak{l}_{(\mathfrak{h},0)}=\{x+\xi\mid x\in\mathfrak{h},\,\xi|_{\mathfrak{h}}=0\}$, which is the conormal subalgebra to $\mathfrak{h}\subset\mathfrak{g}$. This is the infinitesimal analogue of the conormal Lagrangian $N^*[n](\mathrm{B}H\to\mathrm{B}G)$ appearing in Conjecture~\ref{conj:main}. The twisted case $\psi\neq 0$ is parallel to the 2-cochain twist in the finite-group SymTFT; however, as we discussed in the end of §\ref{sec:classical_continuous_SymTFT}, to integrate the twist to a gerbe on $\mathrm{B}H$ we need to decorate the classical TFTs with prequantum data.

The two extreme cases recover the Dirichlet and Neumann boundaries:
\begin{itemize}[nosep]
  \item $\mathfrak{h}=0$: then $\mathfrak{l}_{(0,0)}=\mathfrak{g}^\vee$ is the \emph{Dirichlet boundary};
  \item $\mathfrak{h}=\mathfrak{g}$, $\psi=0$: then $\mathfrak{l}_{(\mathfrak{g},0)}=\mathfrak{g}$ is the \emph{Neumann boundary}.
\end{itemize}

For any \emph{semisimple} Lie subalgebra $\mathfrak{h}$, Whitehead's second lemma gives
\begin{equation}
  \mathrm{H}^2_{\mathrm{CE}}(\mathfrak{h};\mathbb{C}) \;\simeq\; (\Lambda^2\mathfrak{h}^\vee)^{\mathfrak{h}} \;=\; 0,
\end{equation}
so the only topological boundary is $\mathfrak{l}_{(\mathfrak{h},0)}$. For non-semisimple $\mathfrak{h}$, non-trivial Chevalley--Eilenberg 2-cocycles can appear: the simplest example is 2-dimensional non-abelian subalgebra, even if we require the ambient Lie algebra $\mathfrak{g}$ to be reductive \cite{EvensLu01,EvensLu02}.

The correspondence in Proposition~\ref{prop:lag-subalg-cotangent} is the most complete result available for the cotangent double, and it is the Lie-algebraic shadow of the discrete parametrization by pairs $(H,\psi)$ in the finite-group SymTFT. However, the full classification of all Lagrangian Lie subalgebras of $D(\mathfrak{g})$ up to the action of $G$ is more subtle: Evens--Li showed that, for $\mathfrak{g}$ complex semisimple, the closed $G\ltimes\mathfrak{g}^\vee$-orbits in the variety $\mathcal{L}(D(\mathfrak{g}))$ of Lagrangian subalgebras are in bijection with the abelian ideals of a fixed Borel subalgebra, giving $2^{\mathrm{rank}(\mathfrak{g})}$ such orbits \cite{EvensLi21}.

\subsection{The Factorizable Double and Belavin--Drinfeld Data}

If $\mathfrak{g}$ is a complex semisimple Lie algebra equipped with a factorizable \emph{quasitriangular} structure (i.e., a classical $r$-matrix, see \cite{Safronov17}), instead of the Drinfeld double, we can also consider the \emph{factorizable double}
\begin{equation}
  \mathfrak{d} \;=\; \mathfrak{g}\oplus\mathfrak{g}, \qquad \langle(x_1,x_2),(y_1,y_2)\rangle \;=\; \langle x_1,y_1\rangle - \langle x_2,y_2\rangle,
\end{equation}
where $\langle-,-\rangle$ is an $\mathrm{ad}$-invariant inner product on $\mathfrak{g}$. The diagonal $\mathfrak{g}_\Delta\subset\mathfrak{g}\oplus\mathfrak{g}$ is always Lagrangian, and the problem of classifying all other Lagrangian subalgebras is substantially richer than in the cotangent case.

The first complete answer in this setting is due to Karolinsky \cite{Karolinsky99} and Delorme \cite{Delorme01}, building on the classification of classical $r$-matrices by Belavin--Drinfeld \cite{BelavinDrinfeld82}. A \emph{generalized Belavin--Drinfeld triple} $(S,T,d)$ consists of two subsets $S,T$ of the simple roots and a root-system isomorphism $d\colon\langle S\rangle\xrightarrow{\sim}\langle T\rangle$. Given such a triple together with a Lagrangian subspace $V\subseteq\mathfrak{z}_S\oplus\mathfrak{z}_T$ of the center of the corresponding Levi factors, Delorme constructs a standard Lagrangian subalgebra
\begin{equation}
  \mathfrak{l}_{(S,T,d,V)} \;\subset\; \mathfrak{g}\oplus\mathfrak{g}.
\end{equation}
The classification theorem states that every $(G\times G)$-orbit in $\mathcal{L}(\mathfrak{g}\oplus\mathfrak{g})$, the variety of Lagrangian subalgebras, contains exactly one such standard representative; the normalizer of $\mathfrak{l}_{(S,T,d,V)}$ in $G\times G$ is an explicit subgroup $R_{S,T,d}$, and the orbit is isomorphic to $(G\times G)/R_{S,T,d}$ \cite{EvensLu02}.

\begin{remark}
Evens--Lu showed that the variety $\mathcal{L}(\mathfrak{g}\oplus\mathfrak{g})$ carries a natural Poisson structure, its irreducible components are smooth, and the $(G\times G)$-orbit closures are smooth spherical varieties \cite{EvensLu01,EvensLu02}. This geometric structure has no analogue in the cotangent case, reflecting the fact that the factorizable double is genuinely richer than the cotangent double. The Poisson geometry of $\mathcal{L}(\mathfrak{g}\oplus\mathfrak{g})$ and its regular orbit partitions were further developed by Lu--Yakimov \cite{LuYakimov09}.
\end{remark}

\begin{remark}[Physical interpretation]
From the field-theoretic side, the classification of Lagrangian Lie subalgebras of a factorizable double can be understood via the folding trick: a topological defect between two Chern--Simons theories with gauge algebra $\mathfrak{g}$ is equivalent to a topological boundary condition for the doubled algebra $\mathfrak{g}\oplus\mathfrak{g}$, hence to a Lagrangian Lie subalgebra of the factorizable double \cite{ArvanitakisColeDemulder25}. The Belavin--Drinfeld data governing the defects are precisely the data classifying such subalgebras.
\end{remark}

Let us close this section by interpreting the classification of Lagrangian subalgebras of a factorizable double in the language of SymTFT. Up to this point, our discussion has proceeded in the usual direction: starting from symmetry data on a physical boundary, we have identified the corresponding bulk SymTFT. One of the conceptual advantages of the SymTFT viewpoint is that this dictionary can also be read in the reverse direction. Namely, once a bulk topological field theory is fixed, its admissible topological boundary conditions may themselves be regarded as possible realizations of generalized symmetry data. In passing to this broader perspective, one should not expect the resulting symmetry to remain invertible. Rather, the point is that the symmetry data are encoded holographically by the system of topological boundaries and defects of the bulk theory. In this sense, anomalous symmetries form a distinguished subclass of generalized symmetries, rather than exhausting the general picture.

From this perspective, factorizable doubles and their Lagrangian subalgebras provide a basic class of examples of this reverse dictionary in classical continuous SymTFT. Ordinary 3-dimensional BF theory corresponds to the cotangent double $D(\mathfrak{g})$, while replacing the trivial cobracket by a nontrivial Lie bialgebra structure deforms the theory to the Drinfeld double of the Chern--Simons theory. In the factorizable case, this double is equivalent, as a quadratic Lie algebra, to $\mathfrak{g} \oplus \mathfrak{g}$, and the classification of its Lagrangian subalgebras becomes part of the classification of possible topological boundary conditions of the bulk theory.

This phenomenon is intrinsically 3-dimensional in its most direct form: it relies on the identification of Chern--Simons theory for a quadratic Lie algebra as a 3-dimensional AKSZ theory. There is therefore no simple higher-dimensional analogue. It is also not an example of the prequantum twists discussed in §\ref{sec:classical_continuous_SymTFT}, which were introduced by analogy with 't Hooft anomalies of finite symmetries. The Lie-bialgebra twist of BF theory changes the quadratic Lie algebra governing the classical bulk theory; it is not merely the addition of a higher cocycle at the prequantum level. It would be interesting to make the SymTFT meaning of these factorizable doubles and their Lagrangian boundary conditions more explicit, and to compare this perspective with the standard classification of classical $r$-matrices, including the Belavin--Drinfeld classification in the semisimple case.

\appendix

\section{A Concrete Model for the Chern--Simons and BF Target}

The derived-geometric definition of $\mathrm{B}G$ and its $2$-shifted symplectic structure is conceptually clean but rather abstract. For applications to topological field theory, it is useful to have a differential-geometric model in which the relevant closed shifted forms can be written explicitly. A concrete model is provided by the nerve $N_\bullet G$ of the delooping Lie groupoid $G\rightrightarrows *$, viewed as a simplicial manifold.

For a Lie group $G$ whose Lie algebra $\mathfrak{g}$ is equipped with an invariant nondegenerate symmetric pairing $\langle-,-\rangle$, Cueca and Zhu construct an explicit $2$-shifted symplectic Lie groupoid model for $(\mathrm{B}G, \omega_k)$ \cite{CuecaZhu23}. More precisely, they show that the nerve
\[
  N_\bullet G = \left\{\cdots\; G\times G \;\mathbin{\substack{\textstyle\rightarrow\\[-0.4ex]\textstyle\rightarrow\\[-0.4ex]\textstyle\rightarrow}}\; G \;\rightrightarrows\; \pt \right\}
\]
carries a $2$-shifted symplectic form $\Omega_\bullet = \Omega - \Theta + 0$, where $\Omega \in \Omega^2(G\times G)$ is the Brylinski--Weinstein form and $\Theta \in \Omega^3(G)$ is the Cartan $3$-form. This realizes $\omega_k$ concretely in the normalized simplicial de Rham double complex, and hence gives an explicit model for the Chern--Simons target $(\mathrm{B}G, \omega_k)$.

As we introduced in §\ref{sec:background-aksz}, the abstract AKSZ target for 3-dimensional BF theory is the 2-shifted cotangent stack $\Tcot{2}{\mathrm{B}G}$, whose 2-shifted symplectic structure is established by Calaque \cite{Calaque19}. For explicit computations it is useful to have a differential-geometric model for this object in the language of Lie groupoids.

Such a model is developed in the PhD thesis of Stefano Ronchi \cite{Ronchi2025}, building on the $n$-dual formalism of Ronchi--Zhu \cite{RonchiZhu2024}. Given a Lie 2-groupoid $\mathcal{G}$, Ronchi applies a 2-dual construction to the tangent VB 2-groupoid of $\mathcal{G}$, producing a 2-shifted cotangent VB 2-groupoid $\Tcot{2}{\mathcal{G}}$. This object carries a tautological 1-form $\vartheta$, and $\omega_{\mathcal{G}} \coloneqq -\mathrm{d}\vartheta$ is closed in the normalized simplicial de Rham complex, yielding a 2-shifted symplectic structure on $\Tcot{2}{\mathcal{G}}$.

For the delooping groupoid $G \rightrightarrows *$, whose nerve is $N_\bullet G$, Ronchi shows that $\Tcot{2}{N_\bullet G}$ is symplectically Morita equivalent to the Artin--Mazur bar construction applied to the ordinary cotangent groupoid $T^*\mathcal{G}$ \cite{CosteDazordWeinstein1987}. Since the cotangent groupoid of $G \rightrightarrows *$ is identified, up to the usual convention, with a coadjoint action groupoid $G \ltimes \mathfrak{g}^\vee \rightrightarrows \mathfrak{g}^\vee$, this gives a concrete simplicial Lie-groupoid model for the 2-shifted cotangent of $\mathrm{B}G$.

\begin{remark}
The precise identification of Ronchi's Lie-groupoid model with Calaque's derived stack $\Tcot{2}{\mathrm{B}G}$ at the level of derived algebraic geometry has not been established; Ronchi explicitly leaves this comparison as future work \cite{Ronchi2025}.
\end{remark}

\section{Topological Boundaries and Domain Walls}

To formulate topological boundary conditions and domain walls in the explicit simplicial model for $\Tcot{n}{\mathrm{B}G}$, one needs not only the shifted symplectic form but also a theory of morphisms and higher morphisms between Lie $\infty$-groupoids compatible with Morita equivalence. We recall the relevant definitions and sketch the expected higher categorical structure.

\begin{definition}
  A \emph{simplicial manifold} is a functor $M_\bullet \colon \Delta^{\mathrm{op}} \to \operatorname{Diff}$. It satisfies the \emph{Kan condition} if for every $n \geq 1$ and every $0 \leq k \leq n$, the horn map
  \[
    (d_0,\ldots,\widehat{d_k},\ldots,d_n) \colon M_n \longrightarrow \Lambda^k_n(M_\bullet)
  \]
  is a surjective submersion, where $\Lambda^k_n(M_\bullet)$ is the $k$-th horn: the subobject of $M_{n-1}^{\times n}$ consisting of tuples $(x_0,\ldots,\widehat{x_k},\ldots,x_n)$ with $d_i x_j = d_{j-1} x_i$ for all $i < j$ with $i,j \neq k$. A simplicial manifold satisfying the Kan condition is a \emph{Lie $\infty$-groupoid} \cite{Zhu2009}. A \emph{Lie $n$-groupoid} is a simplicial manifold $M_\bullet$ for which the horn maps are surjective submersions for $1 \leq m \leq n$ and diffeomorphisms for $m > n$ and inner horns $0 < k < m$ (with the outer horn maps for $m > n$ remaining surjective submersions).
\end{definition}

\begin{definition}
  A \emph{hypercover} $f_\bullet \colon M_\bullet \to N_\bullet$ is a morphism of simplicial manifolds such that for every $n \geq 0$ and every $0 \leq k \leq n$, the relative horn map
  \[
    M_n \longrightarrow N_n \times_{\Lambda^k_n(N_\bullet)} \Lambda^k_n(M_\bullet)
  \]
  is a surjective submersion. Two Lie $\infty$-groupoids are \emph{Morita equivalent} if they present the same differentiable stack, or equivalently, if they are connected by a zigzag of hypercovers.
\end{definition}

\begin{definition}
  An \emph{$n$-simplicial manifold} is a functor \[M_{\underbrace{\bullet,\ldots,\bullet}_{n}} \colon (\Delta^{\mathrm{op}})^n \to \operatorname{Diff}.\] A \emph{correspondence} of Lie $\infty$-groupoids from $\mathcal{X}_\bullet$ to $\mathcal{Y}_\bullet$ is a bisimplicial manifold $M_{\bullet,\bullet}$ with $M_{\bullet,0} \simeq \mathcal{X}_\bullet$ and $M_{0,\bullet} \simeq \mathcal{Y}_\bullet$, and such that $M_{\bullet,m}$ and $M_{m,\bullet}$ are Lie $\infty$-groupoids for all $m \geq 0$. More generally, an \emph{$n$-fold correspondence} is an $n$-simplicial manifold satisfying the Kan condition in each simplicial direction.
\end{definition}

These definitions fit together into an $(\infty,n)$-categorical structure of correspondences of Lie $\infty$-groupoids. The objects are Lie $\infty$-groupoids, the $k$-morphisms are $k$-fold correspondences for $1 \leq k \leq n$. Equivalence between two $k$-fold correspondences $M$ and $N$ with the same source and target is not given by a fiberwise diffeomorphism, but by a zigzag of hypercovers
\[
  M \xleftarrow{\;\sim\;} P \xrightarrow{\;\sim\;} N
\]
of $k$-simplicial manifolds compatible with source and target. Equivalences between such zigzags are in turn zigzags of hypercovers between the $(k+1)$-simplicial manifolds, and so on at level $k > n$.

\begin{remark}
  The rigorous construction of this $(\infty,n)$-category of correspondences of Lie $\infty$-groupoids, and in particular the higher generalized morphisms and Morita localisation that make the mapping spaces well-defined, is the subject of ongoing work of Blohmann, Krishna, and Zhu \cite{BlohmannKrishnaZhu}. Their announced framework — relating principal groupoid bibundles, anafunctors, and classifying maps in the $\infty$-categorical setting — is the natural differential-geometric language in which Lagrangian correspondences between the simplicial models of \S\ref{sec:classical_continuous_SymTFT} should ultimately be formulated.
\end{remark}

To state what a shifted symplectic structure or a Lagrangian means in the simplicial manifold setting, one works with \emph{multiplicative differential forms} on Lie $\infty$-groupoids, as developed in the context of Lie groupoids by Bursztyn--Cabrera--Ortiz \cite{BursztynCabreraOrtiz12} and in the higher setting by Cueca--Zhu \cite{CuecaZhu23}.

\begin{definition}
  Let $M_\bullet$ be a simplicial manifold. A \emph{degree-$p$ simplicial $q$-form} on $M_\bullet$ is an element
  \[
    \omega \in \bigoplus_{k \geq 0} \Omega^{q-k}(M_k),
  \]
  of total degree $p+q$ in the normalized simplicial de Rham double complex $\mathrm{NdR}(M_\bullet)$, whose rows are the de Rham complexes $(\Omega^\bullet(M_k), \mathrm{d})$ and whose columns are the normalized simplicial cochains with differential $\delta = \sum_i (-1)^i d_i^*$. Such a form is \emph{closed} if it is annihilated by the total differential $\mathrm{d} + \delta$.

  A \emph{$p$-shifted symplectic structure} on a Lie $\infty$-groupoid $M_\bullet$ is a closed degree-$(p+2)$ simplicial $2$-form $\omega \in \mathrm{NdR}^{p+2}(M_\bullet)$ that is nondegenerate in the sense that the induced map $TM_0 \to T^*M_0[p]$ on tangent and cotangent complexes of the space of objects is a quasi-isomorphism.
\end{definition}

\begin{definition}
  Let $(M_\bullet, \omega)$ be a $p$-shifted symplectic Lie $\infty$-groupoid. A \emph{Lagrangian} in $(M_\bullet, \omega)$ is a morphism of simplicial manifolds $f_\bullet \colon L_\bullet \to M_\bullet$ together with a nullhomotopy $\lambda$ of the pullback $f_\bullet^*\omega$ in $\mathrm{NdR}(L_\bullet)$, i.e.\ an element $\lambda \in \mathrm{NdR}^{p+1}(L_\bullet)$ with $(\mathrm{d} + \delta)\lambda = f_\bullet^*\omega$, such that the induced map $TL_0 \to (TM_0|_{L_0})^{\perp_\omega}$ is a quasi-isomorphism onto the $\omega$-orthogonal complement. A \emph{Lagrangian correspondence} from $(L_\bullet, \lambda_L)$ to $(R_\bullet, \lambda_R)$ in $(M_\bullet, \omega)$ is a Lagrangian in $(M_\bullet \times \overline{M}_\bullet, \omega \ominus \omega)$ whose restrictions to the two factors recover $L_\bullet$ and $R_\bullet$ respectively, where $\overline{M}_\bullet$ denotes $M_\bullet$ with symplectic form $-\omega$.
\end{definition}

\begin{remark}
  In the case of the nerve $N_\bullet G$ with the $2$-shifted symplectic form of Cueca--Zhu \cite{CuecaZhu23}, the above definitions recover the classical notion of a quasi-symplectic groupoid and its Lagrangian subgroupoids studied by Xu and others \cite{Xu04}. The data $(\lambda, f_\bullet)$ is the simplicial-geometric analogue of the Lagrangian structure in the sense of PTVV \cite{PTVV13} and Calaque \cite{Calaque15}, and the two notions are expected to agree under the conjectural comparison between Lie $\infty$-groupoid models and derived Artin stacks.
\end{remark}

\bibliographystyle{alpha}
\bibliography{reference}

\end{document}